\documentclass[sigconf,screen]{acmart}

\usepackage{todonotes}
\usepackage{svg}
\AtBeginDocument{%
  }

\setcopyright{acmlicensed}
\copyrightyear{2026}
\acmYear{2026}
\acmDOI{10.1145/3803437.3805566}

\acmConference[Conference FSE'26]{ACM International Conference on the Foundations of Software Engineering}{July 05--09,
  2026}{Montréal, QC, Canada}
\acmISBN{978-1-4503-XXXX-X/2018/06}

\begin{document}

\title{Round-trip Engineering for Tactical DDD: A Constraint-Based Vision for the Masses}

\author{Weixing Zhang}
\orcid{0000-0003-2890-6034}
\affiliation{%
  \institution{Karlsruhe Institute of Technology}
  \city{Karlsruhe}
  \country{Germany} 
}
\email{weixing.zhang@kit.edu}

\author{Mario Herb}
\orcid{0009-0008-9423-4786}
\affiliation{%
  \institution{esentri AG}
  \city{Karlsruhe}
  \country{Germany}}
\email{mario.herb@esentri.com}

\author{Martin Armbruster}
\orcid{0000-0002-2554-4501}
\affiliation{%
  \institution{Karlsruhe Institute of Technology}
  \city{Karlsruhe}
  \country{Germany}
}
\email{martin.armbruster@kit.edu}

\author{Bowen Jiang}
\orcid{0009-0001-8168-0866}
\affiliation{%
 \institution{Karlsruhe Institute of Technology}
 \city{Karlsruhe}
 \country{Germany}}
 \email{bowen.jiang@kit.edu}

\author{Marcel Vielsack}
\affiliation{%
  \institution{esentri AG}
  \city{Karlsruhe}
  \country{Germany}}
  \email{marcel.vielsack@esentri.com}

\author{Anne Koziolek}
\orcid{0000-0002-1593-3394}
\affiliation{%
  \institution{Karlsruhe Institute of Technology}
  \city{Karlsruhe}
  \country{Germany}}
\email{koziolek@kit.edu}

\renewcommand{\shortauthors}{Zhang et al.}

\begin{abstract}
Despite Domain-Driven Design's proven value in managing complex business logic, a fundamental semantic expressiveness gap persists between generic modeling languages and tactical DDD patterns, causing continuous divergence between design intent and implementation. We envision a constraint-based tactical modeling environment that transforms abstract architectural principles into explicit, tool-enforced engineering constraints. At its core is a DDD-native metamodel where tactical patterns are first-class modeling primitives, coupled with a real-time constraint verification engine that prevents architectural violations during modeling, and bidirectional synchronization mechanisms that maintain model-code consistency through round-trip engineering. 
This approach aims to democratize tactical DDD by embedding expert-level architectural knowledge directly into modeling constraints, enabling small teams and junior developers to build complex business systems without sacrificing long-term maintainability. 
By lowering the technical barriers to DDD adoption, we envision transforming tactical DDD from an elite practice requiring continuous expert oversight into an accessible engineering discipline with tool-supported verification.
\end{abstract}



\received{22 January 2026}
\received[accepted]{17 March 2026}

\renewcommand{\footnotetextcopyrightpermission}[1]{%
  {\color{red}\textbf{This is a preprint of a paper accepted at FSE 2026 
  (ACM International Conference on the Foundations of Software Engineering). 
  The final version will appear in the ACM Digital Library.}}%
  \par\vspace{0.5em}%
}

\maketitle

\section{Introduction}


Domain-Driven Design (DDD)~\cite{evans2004domain}, since its introduction in 2003, has become an established methodology for managing complex business logic. However, the practice of tactical DDD still faces inadequate tool support. A recent systematic literature review~\cite{ozkan2025domain} that analyzed 36 peer-reviewed studies identified the persistent ``model-code gap'' as a primary technical challenge in DDD adoption: domain models frequently diverge from their code implementations~\cite{zhang2023manual}\cite{zhang2025empirical}, undermining the core promise of DDD to maintain alignment between domain understanding and software structure.


The root cause of this problem lies in the insufficient semantic expressiveness of existing modeling tools. Neither generic UML tools nor code-first frameworks can adequately encode the semantic constraints of DDD patterns, resulting in model validation relying on manual review rather than tool enforcement~\cite{ozkan2025domain}. This tool-level deficiency is a significant contributor to the model-code gap.

We propose a constraint-based round-trip engineering environment that elevates tactical DDD patterns to first-class citizens of the modeling language, rather than auxiliary annotations to UML. In this environment, \emph{Aggregate Root}, \emph{Entity}, and \emph{Value Object} are no longer class diagram stereotypes that must be manually maintained, but modeling primitives with built-in semantic constraints. For example, when modeling an \emph{aggregate}, if a developer attempts to have an \emph{entity} directly reference an \emph{entity} from another \emph{aggregate}—which violates \emph{aggregate} boundary rules—the verification engine will immediately reject this operation at modeling time and suggest using an identifier \emph{Value Object} instead. By encoding expert-level architectural knowledge as tool-enforced constraints, this vision aims to lower the barrier to tactical DDD, enabling small teams and junior developers to build complex business systems without sacrificing long-term maintainability.

\section{Gap Analysis: The Semantic Deficit in DDD Tooling}
The practice of tactical DDD faces a structural dilemma in tool support. Existing solutions can be categorized into several approaches:

\paragraph{Generic UML tools exhibit insufficient semantic expressiveness.} Tools such as Enterprise Architect and Visual Paradigm represent DDD patterns—\emph{Aggregate Root}, \emph{Entity}, \emph{Value Object}—as stereotypes on class diagrams, which are essentially metadata annotations~\cite{ozkan2025domain}. This approach cannot encode the semantic constraints of patterns~\cite{perillo2009daileon, le2018domain}. For example, tools cannot prevent developers from placing \emph{Entity} classes outside their owning \emph{aggregates}, nor can they enforce the immutability of \emph{Value Object}~\cite{maddodi2020aggregate, hippchen2017designing}. Core DDD rules, such as \emph{Aggregate} boundary integrity checks and \emph{Entity} lifecycle dependency validation, cannot be automatically verified by tools at modeling time. 
While tools like JQAssistant~\cite{jqassistant2025} can check some DDD rules at compile time, such checks operate at the code level rather than during modeling, and current implementations cover only a subset of possible constraints. Most violations are still discovered during code review or at runtime.

\paragraph{Pure code-first frameworks lack model views.} Frameworks such as JMolecules~\cite{JMolecules} annotate the DDD roles of classes through marker interfaces or annotations, providing partial constraint checking at compile time. However, such solutions have two fundamental deficiencies: first, the absence of visual models prevents domain experts from participating in modeling discussions—domain experts typically do not read code~\cite{kapferer2020domain}, yet core DDD practices such as 
EventStorming~\cite{brandolini2018introducing} and Context Mapping~\cite{evans2004domain} depend on shared visual 
artifacts; second, model evolution tracking relies on code commit history, lacking a systematic overview of high-level architectural changes such as Bounded Contexts and Aggregate structures~\cite{braun2021tackling}.

\paragraph{Strategic alignment alone is insufficient.} Practitioners have also attempted to reduce DDD's reliance on expert judgment through strategic-level guidance—for example, using bounded contexts to define microservice boundaries~\cite{zhong2024domain}. However, such alignment cannot prevent tactical violations: aggregate boundaries can still be breached and domain invariants compromised within correctly bounded services~\cite{zhong2024domain}.

In summary, existing tools do not treat DDD patterns as first-class citizens of the modeling language - none
is specifically designed for DDD's tactical patterns~\cite{le2019jdomainapp, snoeck2022agile}.

\section{Vision: DDD-Native Round-trip Engineering}

We propose a constraint-based tactical modeling environment realized through four components: 
1) the \textbf{Domain Modeler} provides a graphical modeling interface based on a native DDD metamodel; 2) the \textbf{Domain Model Verifier} performs real-time constraint checking and offers intelligent repair suggestions, recent advances in GenAI also show potential for enhancing automated repair suggestions~\cite{li2024generative}\cite{li2024exploring}; 3) the \textbf{Domain Code Generator} enables model-to-code incremental synchronization; 4) the code-to-model direction leverages the existing \textbf{Domain Mirror} component~\cite{dmirror}. Together, these transform DDD semantic rules from implicit expert knowledge into tool-enforced constraints.
The Domain Modeler provides a graphical modeling interface with real-time visual feedback, aligned with the vision of metamodel-based language workbenches~\cite{zhang2023rapid}.

\subsection{Metamodel Hierarchy}
In contrast to mapping DDD patterns to generic UML metaclasses, we propose a native DDD metamodel in which the containment relationships and dependency rules among tactical patterns are explicitly encoded as metamodel hierarchies. Specifically: \emph{Aggregate} is the top-level modeling unit that maintains internal consistency through \emph{AggregateRoot} as the sole entry point, explicitly containing \emph{Entity} and \emph{ValueObject}; \emph{Entity} possesses unique identity and lifecycle, while \emph{ValueObject} defines equality through attributes and is immutable; \emph{DomainEvent} is published by \emph{Aggregate} to capture facts about aggregate state changes. This metamodel structure inherently prohibits cross-aggregate entity references, i.e.,
inter-aggregate associations can only be realized through \emph{ValueObject}-typed identifiers. Additionally, the metamodel includes \emph{Repository} and \emph{DomainService} as architectural connectors: \emph{Repository} abstracts the persistence access of \emph{Aggregates} with a one-to-one correspondence to \emph{Aggregates}; \emph{DomainService} encapsulates domain logic that does not naturally belong to a single \emph{Entity} or \emph{ValueObject}~\cite{evans2004domain}. These elements, together with the core tactical patterns, constitute a complete DDD metamodel that enables constraint verification to cover the entire spectrum from domain modeling to architectural organization.

\subsection{Constraint Verification and Intelligent Guidance}
Based on the above metamodel, we identify two types of DDD semantic constraints: (1) structural constraints that specify legal composition relationships among patterns (e.g., value objects cannot contain entity references); (2) behavioral constraints that specify the operational semantics and state evolution rules of patterns, to the extent decidable by static analysis (e.g., aggregate root methods are the sole entry point for modifying internal aggregate state). The verification engine checks these constraints in real-time during modeling operations and provides contextualized repair suggestions. 
For example, when a developer attempts to define a public setter method on an \emph{Entity} that allows external modification of aggregate state bypassing the \emph{AggregateRoot}, the behavioral constraint verifier flags this violation and suggests encapsulating the state change within an \emph{AggregateRoot} method.
This tool-level intelligent guidance embeds DDD expert architectural decision knowledge into the modeling workflow, enabling junior developers to naturally acquire tactical DDD best practices.
Unlike existing constraint-based approaches that focus on post-hoc validation or pattern-based detection of violations~\cite{herwigpattern}, our approach embeds tactical DDD constraints directly into both the modeling language's metamodel and the tool's verification engine, supporting continuous validation during model evolution.
While our current repair suggestions are rule-based, recent work on LLM-based co-evolution of DSL definitions and instances~\cite{zhang2026leveraging} suggests potential for more adaptive, context-aware guidance in future iterations.

\subsection{Bidirectional Synchronization Mechanism}

To bridge the gap between model and code, we adopt a bidirectional synchronization strategy.
In the code-to-model direction, we leverage the existing Domain Mirror component~\cite{dmirror}, which uses reflection (specifically the Java Reflection API) to transfer meta-information from code to the model level. In the model-to-code direction, the Domain Code Generator to be developed will be inspired by the Vitruvius framework~\cite{klare2021enabling}, a view-based approach for consistency preservation in model-driven development, 
to maintain view consistency through incremental model transformations. 
Specifically, model changes will be captured by a delta domain metamodel and translated into corresponding code modifications through declarative transformation rules.

\section{Key Research Challenges}
Realizing the above vision requires systematically addressing the following core challenges:
\subsection{Challenge 1: Balancing Completeness and Practicality of DDD Constraint Rules}
Tactical DDD patterns contain well-defined semantic constraints, but there is currently a lack of systematic research on how to translate these constraints into tool-level verification rules. DDD's core rules such as ``entities cannot reference across aggregate boundaries'' and ``value objects must be immutable'' can be enforced as hard constraints, however, DDD practice requires consideration of additional factors, including design recommendations proposed by the DDD community (such as Vernon's ``design small aggregates'' principle~\cite{vernon2013implementing}) and engineering trade-offs (such as performance optimization and legacy compatibility) that depend on specific contexts.
The challenge is designing a tool that ensures semantic rigor without becoming an obstacle to real-world projects. One pragmatic approach is to allow developers to override specific constraints with documented justifications, or to disable certain constraints for specific projects.

\subsection{Challenge 2: Incremental Delta Identification and Propagation}
\label{sec:chal_2}
The primary challenge of incremental synchronization is identifying model changes. When developers add a new value object attribute to an aggregate,
the system must distinguish between an incremental modification and a scenario requiring complete code regeneration. 
A naive approach would be full regeneration, but this 
risks of overwriting manually added business logic.
DDD patterns make this harder: 
dependencies exist among patterns, and changes to a single element may violate aggregate boundaries or consistency rules.
To address this and in building on the mapping of software architecture models to source code using Vitruvius~\cite{mazkatli2025continuous}, we plan to define a delta metamodel that explicitly expresses the semantics of model changes, coupled with declarative transformation rules to enable precise mapping from models to code.

\subsection{Challenge 3: Round-trip Implementation}
Round-trip engineering includes two directions: the model-to-code direction (discussed in Section~\ref{sec:chal_2}) and the code-to-model direction, which is realized through the existing Domain Mirror component~\cite{dmirror} that uses Java Reflection API to transfer code structure information to the model level. The key challenge is the consistency guarantee between the two directions, i.e., the structure assumptions when the Domain Code Generator generates code must match the identification logic when the Domain Mirror does reflection. For example, when Domain Code Generator generates an entity class, it assumes the class contains an identifier field, and Domain Mirror must also be able to recognize the correspondence between this field and the entity identifier in the model when doing reflection; if developers manually rename or delete this field, the round-trip synchronization will cause inconsistency between the model and code. 
This consistency challenge extends beyond DDD-specific tooling; maintaining coherent relationships among software artifacts, running systems, and models is a broader concern in continuous software development~\cite{armbruster2025process}.

\begin{figure*}[tb]
    \centering
    \includegraphics[width=\textwidth]{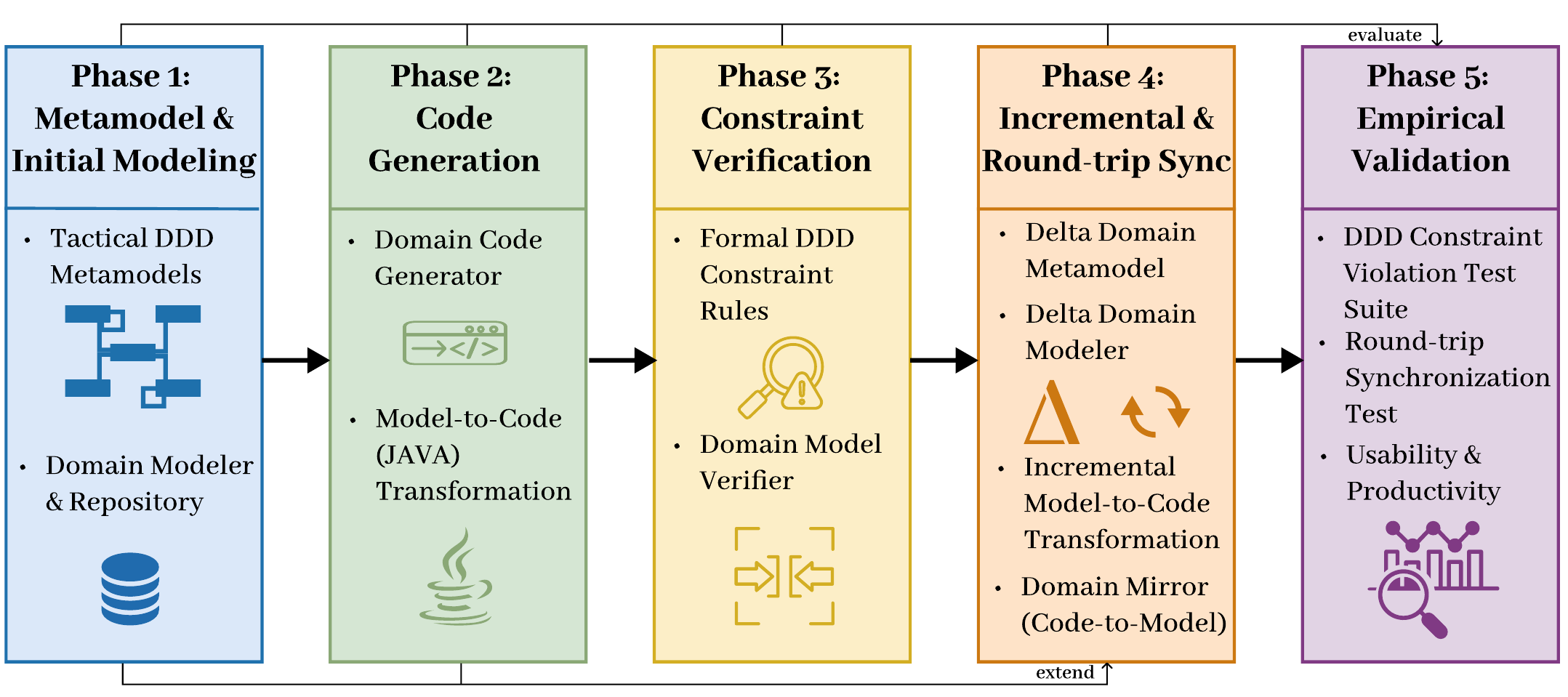}
    \caption{Overview of Research Roadmap}
    \label{fig:roadmap}
\end{figure*}

\section{Research Roadmap}
Based on the components and challenges mentioned above, our Research Roadmap will include the following five phases (Figure~\ref{fig:roadmap}):

\subsection{Phase 1: Metamodel and Initial Modeling}
We will formalize tactical DDD elements into a metamodel that accurately represents core concepts such as Aggregates, Entities, and Value Objects. Based on this metamodel, we will develop the Domain Modeler component with its visualization capabilities, along with a repository for persisting domain models.

\subsection{Phase 2: Code Generation}
We will develop the initial Domain Code Generator component, which enables the generation of a source code representation of a given domain model. Building on the Vitruvius framework~\cite{klare2021enabling}, 
we primarily target Java, with other languages to be explored and developed.

\subsection{Phase 3: Constraint Verification}
Based on existing literature and a literature search, we will identify and formally define the DDD constraint rules, addressing challenge 1 in the process. The resulting constraints will be included in the developed Domain Model Verifier component, which checks the constraints and suggests possibilities for repairing constraint violations.

\subsection{Phase 4: Incremental and Round-trip Synchronization}
We will enhance the Domain Modeler and Domain Code Generator with incremental and round-trip synchronization, addressing challenges 2 and 3. This involves specifying a delta domain metamodel to express DDD-typical changes, enabling incremental model-to-code transformations that modify only affected code portions. The code-to-model direction leverages the existing Domain Mirror component~\cite{dmirror}.

\subsection{Phase 5: Empirical Validation}
Finally, we plan to empirically evaluate and validate the developed components, as outlined in the next section.

\section{Evaluation Strategy}
We plan a three-layered evaluation strategy for validation.

\paragraph{Constraint mechanism effectiveness.} We construct a test suite containing DDD constraint violations identified in~\cite{ozkan2025domain}, such as cross-aggregate entity references, value object mutability violations, and aggregate boundary leaks. We measure constraint rule coverage, precision/recall rates, and whether repair suggestions are semantically correct (verified through expert review). 

\paragraph{Round-trip synchronization quality.} Controlled experiments will test whether incremental synchronization preserves semantic equivalence after model changes, prevents overwriting of manually written business logic, and maintains model-code consistency through multiple round-trips. 

\paragraph{Usability and productivity.} We conduct comparative studies where junior and senior developers complete the same modeling tasks using traditional UML tools versus the constraint-based DDD modeling environment. We evaluate modeling time, task quality (assessed through expert review), and the rate of architectural violations in the resulting models. The hypothesis is that constraint guidance will bring junior developers' modeling quality closer to the senior level.
Additionally, preliminary case studies with industry partners will assess whether the tool fits into real development workflows and whether teams find the constraint enforcement helpful or restrictive.

\section{Discussion: Impact and Limitations}
Our Vision will have an impact on two levels.
First, at the engineering practice and methodology level, the constraint-driven modeling environment can lower the entry barrier to tactical DDD—junior developers do not need to master all architectural principles of classic tactical DDD fully, but can avoid deviating from tactical DDD rules through tool-enforced constraints. 
This enables small teams to build complex business systems without relying on continuous expert review, essentially transforming DDD from an experience-driven practice into a tool-verifiable engineering discipline. 
Second, at the software engineering research level, this research will provide an empirical case for constraint-driven domain-specific modeling, with methodological relevance to model-driven engineering and domain-specific language research.

This vision has four limitations. First, it focuses on the Java ecosystem and may not apply to other languages. Second, the vision focuses on tactical DDD patterns and does not cover strategic-level bounded context modeling. Third, the tool enforces DDD rules but does not teach DDD concepts. Teams still need a basic understanding of domain modeling.
Fourth, practical adoption depends on seamless integration between modeling and coding environments; if developers perceive friction, they may bypass the modeling layer entirely.

\bibliographystyle{ACM-Reference-Format}
\bibliography{main}

\appendix

\end{document}